\documentclass[conference,twoside]{IEEEtran}
\usepackage[pass,paperwidth=8.5in,paperheight=11in]{geometry}
\usepackage[dvips]{epsfig}
\usepackage{cite, algorithm, algorithmic ,amsmath, amssymb, amsthm, empheq}
\usepackage[dvips]{graphicx}
\usepackage{subfigure,amsfonts,balance}
\usepackage{epstopdf}

\DeclareMathOperator{\tr}{\mathrm{tr}}
\DeclareMathOperator{\rank}{\mathrm{rank}}
\DeclareMathOperator{\E}{\mathbb{E}}

\theoremstyle{remark}
\newtheorem{remark}{Remark}

\begin{document}
\title{Interference Alignment Designs \\for Secure Multiuser MIMO Systems: \\Rank Constrained Rank Minimization Approach}
\author{
\IEEEauthorblockN{Tung T. Vu$^{1,2}$, Ha Hoang Kha$^2$, Trung Q. Duong$^3$}
\IEEEauthorblockA{\small
$^1$ Duy Tan University, Vietnam (e-mail: vuthanhtung1@dtu.edu.vn) \\
$^2$ Ho Chi Minh City University of Technology, Vietnam (e-mail: hhkha@hcmut.edu.vn) \\
$^3$ Queen's University Belfast, UK (e-mail: trung.q.duong@qub.ac.uk)
    }
        \thanks{}
    \thanks{}
}
\maketitle

\begin{abstract}
In this paper, we formulate the interference alignment (IA) problem for a multiuser multiple-input multiple-output (MIMO) system in the presence of an eavesdropper as a rank constrained rank minimization (RCRM) problem. The aim of the proposed rank minimization IA schemes is to find the precoding and receiver subspace matrices to align interference and wiretapped signals into the lowest dimension subspaces while keeping the desired signal subspace spanning full available spatial dimensions. To deal with the nonconvexity of the rank function, we present two convex relaxations of the RCRM problem, namely nuclear norm (NN) and reweighted nuclear norm (RNN), and transform the rank constraints to equivalent and tractable ones. We then derive a coordinate decent approach to obtain the solutions for IA schemes. The simulation results show that our proposed IA designs outperform the conventional IA design in terms of average secrecy sum rate. On the other hand, our proposed designs perform the same or better than other secure IA schemes which account for low interference and wiretapped signal power rather than for low dimensions of interference and wiretapped signal matrices in the systems which achieve the perfect IA.
\end{abstract}
\IEEEpeerreviewmaketitle
\section{Introduction}
\label{sec:introd}
 Physical layer security (PLS) has recently attracted considerable attention due to its potential to improve the security of communication systems \cite{mukherjee14}. Different from traditional cryptographic methods which are implemented in the network layer, PLS exploits physical characteristics of wireless channels to provide secrecy. Secure PLS approaches have been developed for various communication scenarios, such as multiuser multiple-input multiple-output (MIMO) systems  \cite{Hanif14,Yang14,Geraci12}, relay networks, cognitive radio systems and other networks (see \cite{Wang14,Jeon14,Pei10} and references therein). The transceiver designs to maximize the secrecy rate in the multiuser MIMO interference channels are of great concern. However, such design problems appear mathematically intractable due to their high nonlinearity and nonconvexity \cite{Hanif14,Yang14}. Alternatively, interference alignment (IA) is one of potential techniques to increase the secrecy sum rate (SSR) in  multiuser MIMO systems \cite{Koyl11,Sasaki12,Tung15}. It is proven in \cite{Koyl11} that it is possible for each user in networks to achieve a nonzero secure degrees of freedom (DoF) when using an IA scheme to design precoding matrices at each transmitter (Tx). The secure transmission has been shown to be feasible where the number of antennas at legitimate Tx and receiver (Rx) is greater than those of the eavesdropper \cite{Sasaki12}. The key idea of a secure IA design is to keep the desired signal free from interference and to offer a secure from the eavesdropper. To obtain interference-free desired signal, interferences at each receiver are aligned into a reduced-dimensional receive subspace, and to prevent desired signals from the eavesdropper, the wiretapped signals at the eavesdropper can be aligned into a proper subspace where their powers are minimized, or can be forced to zero \cite{Tung15}.

The underlying secure IA problems are NP-hard and intractable. To handle these mathematical challenges, reference \cite{Tung15} introduced secure IA approaches which aim at minimizing interference and wiretapped signal power rather than reducing the dimension of interference and wiretapped signal subspaces. Recently, for the multiuser MIMO systems not in the security context, instead of minimizing the power of the interference signal, references \cite{Papailiopoulos12,Du13} proposed rank constrained rank minimization (RCRM) problems which ensure that the interferences fall into a low-dimensional subspace and the desired signal spaces span all available spatial dimensions. Motivated by these works, for the multiuser MIMO systems in the presence of an eavesdropper, we adopt the RCRM framework in which the ranks of the wiretapped signal matrices and the subspace spanned by interference signals are minimized subject to full-rank desired signal space. In order to solve the NP-hard nonconvex rank minimization problem, we introduce nuclear norm (NN) and reweighted nuclear norm (RNN) as two convex approximations of the rank function, then propose heuristic IA algorithms to obtain near optimal solutions. Our experimental evaluations reveal that the proposed IA designs outperform the conventional IA design \cite{peters09}, and provide the same or better SSR performance than secure IA designs in \cite{Tung15} for the system in which the perfect IA can be achieved.

\emph{Notations}: Bold lowercase and uppercase letters represent vectors and matrices respectively. $\pmb{X}^H$ denotes the conjugate transposition of matrix $\pmb{X}$. $\pmb{X}\succeq 0$ represents the Hermitian positive semi-definite matrix $\pmb{X}$. $\pmb{I}_d$ and $\pmb{0}$ are respectively an identity matrix with $d$ dimensions and a zero matrix with the appropriate dimensions. $\tr (.)$, $\rank (.)$, $\E(.)$ and $[x]^{+}$  are the trace, rank, expectation and $\max(x,0)$ operators, respectively. $||\pmb{X}||_F$ is the Frobenius norm.  A complex Gaussian random vector with means $\bar{\pmb{x}}$ and covariance $\pmb{R}_{\pmb{x}}$ is represented by $\pmb{x}\sim \mathcal{CN}(\bar{\pmb{x}},\pmb{R}_{\pmb{x}})$. $\sigma_{min}(\pmb{X})$ and $\sigma_i(\pmb{X})$ are respectively the minimum and the $i$-th largest singular value of matrix $\pmb{X}$ while $||\pmb{X}||_{*}=\sum \limits_{i=1}^{\mathrm{rank}\{X\}}\sigma_i(\pmb{X})$ is the nuclear norm of matrix $\pmb{X}$.

\section{System Model and Problem Formulation}
\label{sec:Model}
\begin{figure}[htb!]
\begin{center}
\epsfxsize=6cm
\leavevmode\epsfbox{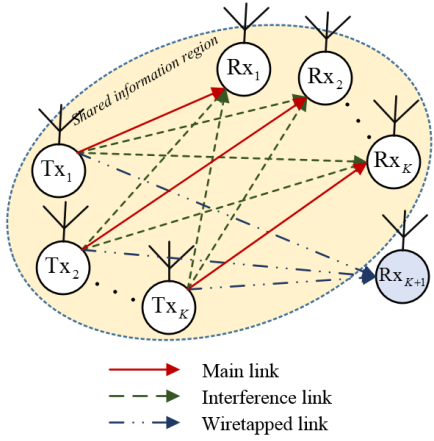}
\vspace{-0.3cm}
\caption{A system model of secure multiuser MIMO communication network.}
\label{fig:DownlinkMIMOIC}
\end{center}
\end{figure}
Consider a MIMO interference channel with $K$ Tx-Rx pairs and an eavesdropper as depicted in Fig. \ref{fig:DownlinkMIMOIC}. Without loss of generality, assume that the eavesdropper is the $(K+1)$-th Rx with the $N_{r_e}$ antennas, each Txs or Rxs is equipped with the same number of antennas $N_t$ and $N_r$, respectively. The $k$-th Tx sends $d$ data streams to the $k$-th Rx. Such a system is denoted by an $(N_t\times N_r,N_{r_e},d)^K$ system \cite{Tung15}. Assuming that $\pmb{H}_{k,\ell} \in \mathbb{C}^{N_r \times N_t}$ is the static flat-fading MIMO channels matrix from the $\ell$-th Tx to the $k$-th Rx, the received signal $\pmb{y}_k \in \mathbb{C}^{N_r \times 1}$ at the $k$-th Rx for $k \in \mathcal{K}=\{1,...,K+1\}$ is given by
\begin{align}\label{Signal:Rx}
\pmb{y}_k = \sum\limits_{\ell = 1}^K \pmb{H}_{k,\ell}\pmb{F}_\ell\pmb{s}_\ell  + \pmb{n}_k
\end{align}
where $\pmb{F}_k \in \mathbb{C}^{N_t \times d}$ is the precoding matrix applied on the signal vector $\pmb{s}_k \in \mathbb{C}^{d \times 1}$ and $\pmb{n}_k \sim \mathcal{CN}\left( 0,\sigma _k^2\pmb{I}_{N_r} \right)$ is a complex Gaussian noise vector. The signal vector $\pmb{s}_k $ is independently identically distributed (i.i.d.) such that $\E\{ {{\pmb{s}_k}\pmb{s}_k^H} \} = {\pmb{I}_{{d}}}$. We consider the scenario that a communication group involves
the eavesdropper (the $(K+1)$-th Rx) but the information is desired to be secret to him \cite{Tung15}. Thus, it can be assumed that the global channel state information (CSI) is available.

The problem of our IA design for secure multiuser MIMO systems is to find
the precoder matrices $\{\pmb{F}_k\}_{k=1}^K$ and the receiving signal subspace matrix $\{\pmb{W}_k\}_{k=1}^{K}$ which satisfy the three following conditions \cite{Tung15}
\begin{eqnarray}\label{IA:SecrecyConditions1}
  \mathrm{rank}\big(\pmb{W}_k^H\pmb{H}_{k,k}\pmb{F}_k\big) &=& d  \\\label{IA:SecrecyConditions2}
  \pmb{W}_k^H\pmb{H}_{k,\ell}\pmb{F}_\ell &=& \mathbf{0};\,\, \forall \ell \neq k, \ell \in \mathcal{K}\\\label{IA:SecrecyConditions3}
  \pmb{H}_{K+1,\ell}\pmb{F}_\ell &=& \mathbf{0}; \,\, \forall \ell \in \mathcal{K},
\end{eqnarray}
where \eqref{IA:SecrecyConditions1} enforces the useful signal to span all $d$ dimensions, \eqref{IA:SecrecyConditions2} and \eqref{IA:SecrecyConditions3} ensure all the interference subspaces and wiretapped signal subspace to have zero dimensions.

\begin{remark}
The feasibility of a set of three linear IA equations above has been studied \cite{Tung15}. It is shown in \cite{Tung15} that perfect IA for an $\left(N_t \times N_r, N_{r_e}, d \right)^{K}$ system is almost surely achieved under properness condition $N_t-d\geq N_e$ and $N_r\geq Kd$. If this condition is not satisfied, the system is called improper.
\end{remark}

For ease of presentation, we define the desired signal matrix $\pmb{S}_k \in \mathbb{C}^{d\times d}$, interference matrix $\pmb{J}_k \in \mathbb{C}^{d\times (K-1)d}$ and the wiretapped signal matrix $\pmb{S}_e \in \mathbb{C}^{N_{r_e}\times Kd}$, for all $k \in \mathcal{K}$, respectively as follows
\begin{eqnarray}\label{Sk}
\pmb{S}_k \left(\pmb{W}_k,\pmb{F}_k\right) \triangleq \pmb{W}_k^H\pmb{H}_{k,k}\pmb{F}_k, \\ \label{Jk}
\pmb{J}_k \left(\pmb{W}_k,\left\{\pmb{F}_\ell\right\}_{\ell=1,\ell \neq k}^K\right) \triangleq \pmb{W}_k^H\left[\left\{\pmb{H}_{k,\ell}\pmb{F}_\ell\right\}_{\ell=1,\ell \neq k}^K\right],\\ \label{Se}
\pmb{S}_e\left(\left\{\pmb{F}_k\right\}_{k=1}^K\right) \triangleq \left[\left\{\pmb{H}_{{K+1},\ell}\pmb{F}_\ell\right\}_{\ell=1}^K\right].
\end{eqnarray}
Accordingly, \eqref{IA:SecrecyConditions1}, \eqref{IA:SecrecyConditions2} and \eqref{IA:SecrecyConditions3} can be restated in terms of ranks
\begin{eqnarray}\label{rankSk}
\mathrm{rank}\left(\pmb{S}_k\right) &=& d,\\ \label{rankJk}
\mathrm{rank}\left(\pmb{J}_k\right) &=& 0,\\ \label{rankSe}
\mathrm{rank}\left(\pmb{S}_e\right) &=& 0,
\end{eqnarray}
for all $k \in \mathcal{K}$.
To aim at obtaining $d$ interference-free dimensions for every Rx and zero wiretapped-signal-space dimensions, we seek the proper precoding and receiving subspace matrices to minimize the rank of interference matrices in \eqref{rankJk} and the rank of wiretapped-signal matrix in \eqref{rankSe} subject to the full rank of the desired signal matrix. Thus, the design problem of interest can be mathematically expressed as
\begin{subequations}\label{mainOptPro}
\begin{align}\label{CF}
 \underset{\left\{\pmb{F}_k\right\}_{k=1}^K,\left\{\pmb{W}_k\right\}_{k=1}^{K}}{\text{min}} \qquad
 &\sum\limits_{k=1}^K\mathrm{rank}\left(\pmb{J}_k\right) + \mathrm{rank}\left(\pmb{S}_e\right)\\ \label{rankConstraint}
 \text{s.t.} \,\,\,\,\,\qquad\qquad &\mathrm{rank}\left(\pmb{S}_k\right) = d, \,\,\, \forall \ell \in \mathcal{K}.
\end{align}
\end{subequations}
 Both the cost function and constraints in  \eqref{mainOptPro} are non-convex and, thus, problem \eqref{mainOptPro} is mathematically intractable. Therefore, it is highly desired to find the efficient iterative algorithm to obtain suboptimal solution to \eqref{mainOptPro} instead of the globally optimal solution. Specifically, we introduce convex surrogates for this cost function and the feasible solution set. This results another challenging problem is that, although the surrogate of the cost function is convex in either of the two sets of input matrices, the cost function is no longer convex if it is minimized over both sets at the same time. To handle this problem, we utilize a coordinate descent approach to alternatively minimize the cost function over the transmit and then over the receive subspace matrices iteratively. The constrain $\mathrm{rank}\left(\pmb{S}_k\right) = d$ can be replaced approximated with the following closed convex set,
in the same manner as \cite{Papailiopoulos12}
\begin{align}\label{convexContraintApprox}
\pmb{S}_k \succeq \pmb{0},\,\,\, \sigma_{min}(\pmb{S}_k) \geq \epsilon,
\end{align}
where $\epsilon > 0$. Note that, the orthogonality constraints on the precoding and receiver subspace matrices are omitted in problem \eqref{mainOptPro}. This is because that they can be linearly transformed to be orthogonal by using QR factorization, while the ranks of interference, desired signal and wiretapped signal matrices remain unchanged \cite{Papailiopoulos12}.

\section{RCRM based IA approach for Secure MIMO Interference Channels}

\subsection{NN approximation based IA algorithm}
To deal with the nonconvexity of the rank function, NN $||\pmb{X}||_{*}$ is used as a surrogate for $\mathrm{rank}(\pmb{X})$ \cite{fazel04}. Problem \eqref{mainOptPro} can be recast as follows
\begin{subequations}\label{mainOptPro:approx:NN}
\begin{align}\label{CF:approx:NN}
 \underset{\left\{\pmb{F}_k\right\}_{k=1}^K,\left\{\pmb{W}_k\right\}_{k=1}^{K+1}}{\text{min}} \qquad
 &\sum\limits_{k=1}^K ||\pmb{J}_k||_{*} + ||\pmb{S}_e||_{*}\\ \label{rankConstraint:approx:NN}
 \text{s.t.} \,\,\,\,\,\qquad\qquad &\pmb{S}_k \succeq \pmb{0},\,\,\, \sigma_{min}(\pmb{S}_k) \geq \epsilon.
\end{align}
\end{subequations}
Then, we solve the problem above by the coordinate descent approach as follows.
\subsubsection*{Transmit precoder selection}
When $\{\pmb{W}_k\}_{k=1}^{K}$ are fixed, for selecting $\{\pmb{F}_k\}_{k=1}^{K}$, \eqref{mainOptPro:approx:NN} reduces to the optimization problem $\mathcal{P}_{\pmb{F}}^{\mathrm{NN}}$ defined as
\begin{empheq}[box=\fbox]{align}\nonumber
\mathcal{P}_{\pmb{F}}^{\mathrm{NN}}
:
&\underset{\left\{\pmb{F}_k\right\}_{k=1}^K}{\text{min}} \qquad
 \sum\limits_{k=1}^K ||\pmb{J}_k||_{*} + ||\pmb{S}_e||_{*}\\ \label{rankConstraint:approx} \nonumber
& \,\,\,\,\,\,\text{s.t.} \,\,\,\qquad \pmb{S}_k \succeq \pmb{0},\,\,\, \sigma_{min}(\pmb{S}_k) \geq \epsilon.
\end{empheq}
\subsubsection*{Receive subspace selection}
 We use the solution $\left\{\pmb{F}_k\right\}_{k=1}^{K}$ of problem $\mathcal{P}_{\pmb{F}}^{\mathrm{NN}}$ as an input to select $\{\pmb{W}_k\}_{k=1}^{K}$ by solving problems $\mathcal{P}_{\pmb{W}}^{\mathrm{NN}}$ defined as
\begin{empheq}[box=\fbox]{align}\nonumber
\mathcal{P}_{\pmb{W}_k}^{\mathrm{NN}}
:
&\underset{\left\{\pmb{W}_k\right\}_{k=1}^{K}}{\text{min}} \qquad
 \sum\limits_{k=1}^K ||\pmb{J}_k||_{*} \\ \label{rankConstraint:approx} \nonumber
& \,\,\,\,\,\,\text{s.t.} \,\,\,\qquad \pmb{S}_k \succeq \pmb{0},\,\,\, \sigma_{min}(\pmb{S}_k) \geq \epsilon,
\end{empheq}
The optimization problem above is convex and can be efficiently solved using CVX toolbox \cite{cvx}. We then feed the solution $\left\{\pmb{W}_k\right\}_{k=1}^{K}$ of this optimization problem back to $\mathcal{P}_{\pmb{F}}^{\mathrm{NN}}$.
This process is continued over iterations. The step-be-step procedure is stated as Algorithm \ref{alg1}.
\begin{algorithm}[h!]
\caption{: Secure NN IA Algorithm}\label{IterIA}
\begin{algorithmic}[1]\label{alg1}
\STATE Inputs: $d,\pmb{H}_{k,\ell}$,  $\forall k \in \mathcal{K}\cup\{K+1\}$, $\forall \ell \in \mathcal{K}$, $\kappa=0$, $\kappa_{\max }$, where $\kappa$ is the iteration index;
\STATE Initial variables: random matrix $\{\pmb{F}_k^{(0)}\}_{k = 1}^{K}$ satisfied $\pmb{F}_k^{(0)H}\pmb{F}_k^{(0)}=\frac{P_t}{d}\pmb{I}_d$;
\WHILE{$\kappa < \kappa_{\max}$}
\STATE For fixed $\{\pmb{F}_k^{(\kappa)}\}_{k = 1}^K$, select $\{\pmb{W}_k^{(\kappa+1)}\}_{k = 1}^{K}$ by solving $\mathcal{P}_{\pmb{W}_k}^{\mathrm{NN}}$ and orthogonalize $\{\pmb{W}_k^{(\kappa+1)}\}_{k = 1}^{K}$.
\STATE For fixed $\{\pmb{W}_k^{(\kappa+1)}\}_{k = 1}^{K}$, select $\{\pmb{F}_k^{(\kappa+1)}\}_{k = 1}^K$ by solving $\mathcal{P}_{\pmb{F}}^{\mathrm{NN}}$ and orthogonalize $\{\pmb{F}_k^{(\kappa+1)}\}_{k = 1}^K$.
\STATE $\kappa=\kappa+1$;
\STATE Repeat steps 4-6 until convergence or when $\kappa$ reaches the maximum number of iteration $\kappa_{\max}$.
\ENDWHILE
\end{algorithmic}
\end{algorithm}

The NN approximation in the cost function of \eqref{mainOptPro:approx:NN} accounts for the sum of the singular values of the interference matrix and wiretapped-signal matrix, rather than the sum of the number of singular values. Thus, such an approximation may not result in the minimum rank of the cost function of \eqref{mainOptPro}. In order to make a tighter approximation of the rank function, we employ a different surrogate, namely RNN, for the cost function in the following subsection.

\subsection{RNN based IA algorithm}
The RNN approximation of a rank function was provided in the multiuser MIMO interference channels but not in the security context \cite{Du13}. In this paper, we adopt the RNN approach to a secure multiuser MIMO system and develop an iterative RNN algorithm to reduce the wiretapped-signal and interference subspace dimensions. The basic idea of RNN approximation is using the function $\log \det(\pmb{X}+\delta\pmb{I})$, known as a smooth surrogate of $\mathrm{rank}(\pmb{X})$ where $\pmb{X}$ is positive semidefinite matrix. The log-det type function is proved to possibly obtain low-rank solutions to linear matrix inequality problems for positive semidefinite matrices \cite{fazel04}. Using the RNN approximation, the RCRM problem \eqref{mainOptPro} can be rewritten as
\begin{subequations}\label{mainOptPro:approx:RNN}
\begin{align}\label{CF:approx:RNN}
 \underset{\left\{\pmb{F}_k\right\}_{k=1}^K,\left\{\pmb{W}_k\right\}_{k=1}^{K}}{\text{min}}
 &\,\,\,\,\,\qquad\qquad\Omega\\ \label{rankConstraint:approx:RNN}
 \text{s.t.} \,\,\,\,\,\qquad &\pmb{S}_k \succeq \pmb{0},\,\,\, \sigma_{min}(\pmb{S}_k) \geq \epsilon,
\end{align}
\end{subequations}
where $\Omega =
\sum\limits_{k=1}^K\sum\limits_{i=1}^d \log\left(\sigma_i(\pmb{J}_k)+\gamma\right) +
 \sum\limits_{i=1}^{d_e} \log\left(\sigma_i(\pmb{S}_e)+\zeta\right)$,
%
$d_e = \min(N_e,Kd)$, $\gamma$ and $\zeta$ are the arbitrable small positive values to make the approximation resemble the rank function and ensure the stability for the log function.

Since the cost function in \eqref{mainOptPro:approx:RNN} is concave, we apply a majorization-minimization (MM) algorithm to solve problem \eqref{mainOptPro:approx:RNN}. In particular, at the $\kappa$-th iteration, we find the upper bound of the cost function, then, minimize it to ensure that the cost function can reach the optimal minimizer over iterations. The upper bound of the cost function \eqref{CF:approx:RNN} can be obtained by taking the first order Taylor expansions with respect to $\sigma_i(\pmb{J}_k)$ and $\sigma_i(\pmb{S}_e)$ as follows
\begin{align}\label{Jk:upbound}
\log\left(\sigma_i(\pmb{J}_k)+\gamma\right)\leq \log\left(\sigma_i(\pmb{J}_k^{(\kappa)})+\gamma\right)+\frac{\sigma_i(\pmb{J}_k)-\sigma_i(\pmb{J}_k^{(\kappa)})}{\sigma_i(\pmb{J}_k^{(\kappa)})+\gamma}
\end{align}
\begin{align}\label{Se:upbound}
\log\left(\sigma_i(\pmb{S}_e)+\zeta\right)\leq \log\left(\sigma_i(\pmb{S}_e^{(\kappa)})+\zeta\right)+\frac{\sigma_i(\pmb{S}_e)-\sigma_i(\pmb{S}_e^{(\kappa)})}{\sigma_i(\pmb{S}_e^{(\kappa)})+\zeta}
\end{align}
It should be noted that $\pmb{J}_k^{(\kappa)}$, $\pmb{S}_e^{(\kappa)}$ and their singular values are known at the $\kappa$-th iteration. Hence, we iteratively minimize the upper bound given by
\begin{subequations}\label{MinCFUpBound}
\begin{align}\label{CF:Upbound}
 \underset{\left\{\pmb{F}_k\right\}_{k=1}^K,\left\{\pmb{W}_k\right\}_{k=1}^{K}}{\text{min}}
 &\,\,\,\,\,\sum\limits_{k=1}^K\sum\limits_{i=1}^d\frac{\sigma_i(\pmb{J}_k)}{\sigma_i(\pmb{J}_k^{(\kappa)})+\gamma}
 + \sum\limits_{i=1}^{d_e}\frac{\sigma_i(\pmb{S}_e)}{\sigma_i(\pmb{S}_e^{(\kappa)})+\zeta}\\ \label{rankConstraint:approx:RNN}
 \text{s.t.} \,\,\,\,\,\qquad&\qquad\qquad\pmb{S}_k \succeq \pmb{0},\,\,\, \sigma_{min}(\pmb{S}_k) \geq \epsilon,
\end{align}
\end{subequations}
which is rewritten as
\begin{subequations}\label{MinCFUpBound:RNN:1}
\begin{align}\label{CF:JkUpbound:RNN}
 \underset{\left\{\pmb{F}_k\right\}_{k=1}^K,\left\{\pmb{W}_k\right\}_{k=1}^{K}}{\text{min}}
 &\,\,\,\,\,\sum\limits_{k=1}^K||\pmb{\Xi}_k^{(\kappa)}\pmb{J}_k||_{*}
 +||\pmb{\Phi}_e^{(\kappa)}\pmb{S}_e||_{*}\\ \label{rankConstraint:approx:RNN}
 \text{s.t.} \,\,\,\,\,\qquad &\qquad\pmb{S}_k \succeq \pmb{0},\,\,\, \sigma_{min}(\pmb{S}_k) \geq \epsilon.
\end{align}
\end{subequations}
with condition $N_{r_e}< Kd$ or
\begin{subequations}\label{MinCFUpBound:RNN:2}
\begin{align}\label{CF:SeUpbound:RNN}
 \underset{\left\{\pmb{F}_k\right\}_{k=1}^K,\left\{\pmb{W}_k\right\}_{k=1}^{K}}{\text{min}}
 &\,\,\,\,\,\sum\limits_{k=1}^K||\pmb{\Xi}_k^{(\kappa)}\pmb{J}_k||_{*}
 +||\pmb{S}_e\pmb{\Phi}_e^{(\kappa)}||_{*}\\ \label{rankConstraint:approx:RNN}
 \text{s.t.} \,\,\,\,\,\qquad &\qquad\pmb{S}_k \succeq \pmb{0},\,\,\, \sigma_{min}(\pmb{S}_k) \geq \epsilon,
\end{align}
\end{subequations}
with condition $N_{r_e} \geq Kd$, where $\pmb{\Xi}_k^{(\kappa)}\in \mathbb{C}^{d \times d}$ and $\pmb{\Phi}_e^{(\kappa)}\in \mathbb{C}^{{d_e} \times {d_e}}$ are the weight matrices which is defined as
\begin{eqnarray} \label{JkWeighted}
\pmb{\Xi}_k^{(\kappa)} = \pmb{\Psi}_k^{(\kappa)}\pmb{\Upsilon}_k^{(\kappa)}\pmb{\Psi}_k^{(\kappa)H}, \quad
\pmb{\Phi}_e^{(\kappa)} = \pmb{\Delta}_e^{(\kappa)}\pmb{\Theta}_e^{(\kappa)}\pmb{\Delta}_e^{(\kappa)H},
\end{eqnarray}
in which $\pmb{\Psi}_k^{(\kappa)}\in \mathbb{C}^{d \times d}$ are the left singular vectors of $\pmb{J}_k^{(\kappa)}$ , $\pmb{\Delta}_k^{(\kappa)}\in \mathbb{C}^{{d_e} \times {d_e}}$ are the left singular vectors of $\pmb{S}_e^{(\kappa)}$ when $N_{r_e} < Kd$ or the right singular vectors of $\pmb{S}_e^{(\kappa)}$ when $N_{r_e} \geq Kd$, $\pmb{\Upsilon}_k^{(\kappa)}\in \mathbb{C}^{d \times d}$ is the diagonal matrix whose $i$-th diagonal element is equal to $\upsilon_i^{(\kappa)}=\frac{1}{\sigma_i(\pmb{J}_k^{(\kappa)})+\gamma}$ and $\pmb{\Theta}_e^{(\kappa)}\in \mathbb{C}^{{d_e} \times {d_e}}$ is the diagonal matrix whose $i$-th diagonal element is equal to $\theta_i^{(\kappa)}=\frac{1}{\sigma_i(\pmb{S}_e^{(\kappa)})+\gamma}$. The similar proofs of these above transformations can be found in \cite{Du13} and, thus, omitted.

Now, we propose a RNN two-loop algorithm for the RCRM problem \eqref{mainOptPro:approx:RNN}. At the $\kappa$-th iteration, we create an inner loop to find optimal precoding and receive subspace matrices in \eqref{MinCFUpBound:RNN:1} or \eqref{MinCFUpBound:RNN:2} via the coordinated descent approach. The main steps in the $m$-th iteration inside the inner loop of our algorithm are stated in the following.
\subsubsection*{Transmit precoder selection}
By holding $\{\pmb{W}_k\}_{k=1}^{K}$ fixed, we select $\{\pmb{F}_k\}_{k=1}^{K}$ by solving the optimization problem $\mathcal{P}_{\pmb{F}}^{\mathrm{RNN}}$ defined as follows
\begin{empheq}[box=\fbox]{align}\nonumber
\mathcal{P}_{\pmb{F}}^{\mathrm{RNN}}
:
&\underset{\left\{\pmb{F}_k\right\}_{k=1}^K}{\text{min}} \qquad
 \sum\limits_{k=1}^K||\pmb{\Xi}_k^{(\kappa)}\pmb{J}_k||_{*}+ ||\pmb{\Phi}_e^{(\kappa)}\pmb{S}_e||_{*}\\ \label{rankConstraint:approx} \nonumber
& \,\,\,\,\,\,\text{s.t.} \,\,\,\qquad \pmb{S}_k \succeq \pmb{0},\,\,\, \sigma_{min}(\pmb{S}_k) \geq \epsilon.
\end{empheq}
when $N_{r_e}< Kd$, or
\begin{empheq}[box=\fbox]{align}\nonumber
\mathcal{P}_{\pmb{F}}^{\mathrm{RNN}}
:
&\underset{\left\{\pmb{F}_k\right\}_{k=1}^K}{\text{min}} \qquad
 \sum\limits_{k=1}^K||\pmb{\Xi}_k^{(\kappa)}\pmb{J}_k||_{*}+ ||\pmb{S}_e\pmb{\Phi}_e^{(\kappa)}||_{*}\\ \label{rankConstraint:approx} \nonumber
& \,\,\,\,\,\,\text{s.t.} \,\,\,\qquad \pmb{S}_k \succeq \pmb{0},\,\,\, \sigma_{min}(\pmb{S}_k) \geq \epsilon.
\end{empheq}
when $N_{r_e}\geq Kd$.
\subsubsection*{Receive subspace selection}
 We now set the solution $\left\{\pmb{F}_k\right\}_{k=1}^{K}$ of problem $\mathcal{P}_{\pmb{F}}^{\mathrm{RNN}}$ as an input to find receive subspace matrices $\left\{\pmb{W}_k\right\}_{k=1}^{K}$ by solving the two following problem $\mathcal{P}_{\pmb{W}_k}^{\mathrm{RNN}}$ defined as
\begin{empheq}[box=\fbox]{align}\nonumber
\mathcal{P}_{\pmb{W}_k}^{\mathrm{RNN}}
:
&\underset{\left\{\pmb{W}_k\right\}_{k=1}^{K}}{\text{min}} \qquad
 \sum\limits_{k=1}^K||\pmb{\Xi}_k^{(\kappa)}\pmb{J}_k||_{*}\\ \label{rankConstraint:approx} \nonumber
& \,\,\,\,\,\,\text{s.t.} \,\,\,\qquad \pmb{S}_k \succeq \pmb{0},\,\,\, \sigma_{min}(\pmb{S}_k) \geq \epsilon,
\end{empheq}
Each problem above is convex and can be solved by CVX toolbox \cite{cvx}. This process continues until the cost function converges or $m$ attains a specified maximum number of iteration $m_{max}$. We then update input $\pmb{\Xi}_k^{(\kappa)}$ and $\pmb{\Phi}_e^{(\kappa)}$ to the next minimization in $(\kappa+1)$-th iteration. The step-by-step algorithm is described in Algorithm \ref{alg2}.
\begin{algorithm}[h!]
\caption{: Secure RNN IA Algorithm}\label{IterIA}
\begin{algorithmic}[1]\label{alg2}
\STATE Inputs: $d,\pmb{H}_{k,\ell}$,  $\forall k \in \mathcal{K}\cup\{K+1\}$, $\forall \ell \in \mathcal{K}$, $m=\kappa=0$, $\kappa_{\max }$, $m_{max}$;
\STATE Initial variables: random matrix $\{\pmb{F}_k^{(0)}\}_{k = 1}^{K}$ satisfied $\pmb{F}_k^{(0)H}\pmb{F}_k^{(0)}=\frac{P_t}{d}\pmb{I}_d$, $\pmb{\Xi}_k^{(0)}=\pmb{I}_d$ and $\pmb{\Phi}_k^{(0)}=\pmb{I}_d$;
\WHILE{$\kappa < \kappa_{\max}$}
\WHILE{$m < m_{\max}$}
\STATE For fixed $\{\pmb{F}_k^{(m)}\}_{k = 1}^K$, select $\{\pmb{W}_k^{(m+1)}\}_{k = 1}^{K}$ by solving $\mathcal{P}_{\pmb{W}_k}^{\mathrm{RNN}}$ and orthogonalize $\{\pmb{W}_k^{(m+1)}\}_{k = 1}^{K}$;
\STATE For fixed $\{\pmb{W}_k^{(m+1)}\}_{k = 1}^{K}$, select $\{\pmb{F}_k^{(m+1)}\}_{k = 1}^K$ by solving $\mathcal{P}_{\pmb{F}}^{\mathrm{RNN}}$ and orthogonalize $\mathcal{P}_{\pmb{F}}^{\mathrm{RNN}}$ ;
\STATE Update $m=m+1$;
\STATE Repeat steps 5-7 until convergence or when $m$ reaches the maximum number of iteration $m_{\max}$;
\ENDWHILE
\STATE Output: $\{\pmb{F}_k^{(\kappa+1)}\}_{k = 1}^K$ and $\{\pmb{W}_k^{(\kappa+1)}\}_{k = 1}^{K}$;
\STATE Evaluate $\pmb{J}_k^{(\kappa+1)}$ and $\pmb{S}_e^{(\kappa+1)}$ from \eqref{Jk} and \eqref{Se};
\STATE Update $\pmb{\Xi}_k^{(\kappa+1)}$ and $\pmb{\Phi}_e^{(\kappa+1)}$ from \eqref{JkWeighted};
\STATE Update $\kappa=\kappa+1$;
\STATE Repeat steps 4-13 until convergence or when $\kappa$ reaches the maximum number of iteration $\kappa_{\max}$.
\ENDWHILE
\end{algorithmic}
\end{algorithm}

\section{Simulation Results}
\label{sec:Results}
In this experimental evaluation, we run simulation for a $(18\times 12,9,3)^3$ and a $(15\times 15,9,3)^3$ system. Noise variances are normalized $\sigma_k^2=\sigma^2=1$. The Rayleigh fading channel coefficients are generated from the complex Gaussian distribution ${\mathcal{CN}}(0,1)$. We define signal-to-noise-ratio $\text{SNR}=\frac{P_{t}}{\sigma^2}$. In additional to the proposed IA designs, namely NN IA and RNN IA design, two other secure IA algorithms which minimize the power of interference and wiretapped signals are considered, such as the wiretapped signal leakage minimization (WSLM) \cite{Tung15} and zero-forcing wiretapped signal (ZFWS) \cite{Tung15}. It is noted in \cite{Tung15} that these two systems considered are proper for all these four IA methods, i.e., $N_{r_e}\leq \frac{K(N_t+N_r)-(K^2+1)d}{K-1}$, $N_{r_e}\leq N_t-d$ and $N_r \geq Kd$. We present the numerical results averaged over $200$ channel realizations. We then plot and compare the achievable SSR of each IA design and the conventional IA design in which the security context was not taken into account.
The channel capacity at the $k$-th Rx, for $k \in \mathcal{K}$, can be calculated directly from \eqref{Signal:Rx} as \cite{Bazz12}
\begin{align}\label{Cap:Rxk}
\mathcal{R}_k = \log _2\left| \pmb{I}_{N_r} + \pmb{H}_{k,k}\pmb{F}_k\pmb{F}_k^H\pmb{H}_{k,k}^H\pmb{R}_{z_k}^{-1} \right|
\end{align}
where $\pmb{R}_{z_k} = \sum\limits_{\ell = 1,\ell \ne k}^K \pmb{H}_{k,\ell}{\pmb{F}_\ell\pmb{F}_\ell^H\pmb{H}_{k,\ell}^H}  + \sigma _k^2\pmb{I}_{N_r}$ is the interference plus noise correlation matrix in Eq. \eqref{Signal:Rx}. The information leakage rate from the $k$-th Tx to the eavesdropper can computed by
\begin{align}\label{Cap:WiretapRxk}
\mathcal{R}_k^{(e)} = \log _2\left| \pmb{I}_{N_{r_e}} + \pmb{H}_{K+1,k}\pmb{F}_k\pmb{F}_k^H\pmb{H}_{K+1,k}^H\pmb{R}_{e,k}^{-1} \right|
\end{align}
where $\pmb{R}_{e,k} = \sum\limits_{\ell = 1,\ell \ne k}^K \pmb{H}_{K+1,\ell}{\pmb{F}_\ell\pmb{F}_\ell^H\pmb{H}_{K+1,\ell}^H}  + \sigma _k^2\pmb{I}_{N_{r_e}}$ is the interference plus noise correlation matrix at the eavesdropper. The $k$-th Tx-Rx pair can obtain the secrecy rate given by $\mathcal{R}_{S,k}  = [\mathcal{R}_k-\mathcal{R}_k^{(e)}]^{+}$. Hence, the multiuser MIMO system can have the SSR
\begin{align}\label{SSR}
\mathcal{R}_{S} = \sum\limits_{k = 1}^K\mathcal{R}_{S,k} = \sum\limits_{k = 1}^K [\mathcal{R}_k-\mathcal{R}_k^{(e)}]^{+}.
\end{align}
In the following experiments, we set the specific number of iterations for each IA design so that all IA algorithms take comparable times to run  using MATLAB. For each simulation, we run $5$ iterations of the NN IA algorithm, $3$ iterations of the RNN IA algorithm, $50$ iterations of the ZFWS IA algorithm and $25$ iterations of the WLSM IA algorithm. To solve $\mathcal{P}_{\alpha}^{\eta}$ where $\alpha=\{\pmb{F},\pmb{W}_k\}$ and $\eta=\{\mathrm{NN,RNN}\}$, we set $\epsilon=0.1$ and the maximum number of inner loops $m_{max}=3$.

\begin{figure}[htb!]
\begin{center}
\epsfxsize=8cm
\leavevmode\epsfbox{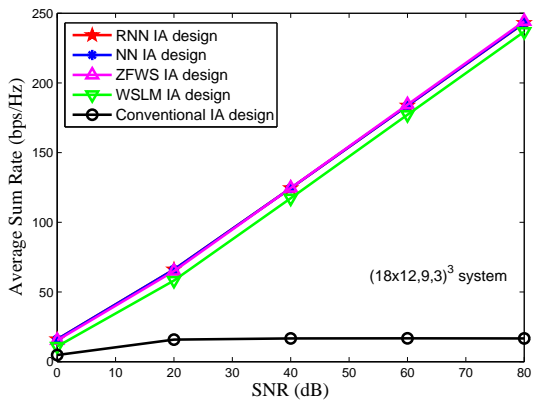}
\vspace{-0.3cm}
\caption{The average SSR versus SNR for ${\left( 18 \times 12,9,3 \right)^3}$ systems.}
\label{fig2}
\end{center}
\end{figure}

Fig. \ref{fig2} plots the average SSR versus $SNR$ for a $(18\times 12,9,3)^3$ system. It can be seen that our proposed IA designs significantly improve the SSR as compared with the conventional IA design. Two proposed designs offer a slight SSR improvement when compared to the WSLM IA scheme while they provide the same SSR performance as ZFWS IA scheme. In this system, our IA algorithms appear to achieve the perfect IA and, thus, the RNN IA approach seems not to show the SSR enhancement as compared with the NN IA approach.

\begin{figure}[htb!]
\begin{center}
\epsfxsize=8cm
\leavevmode\epsfbox{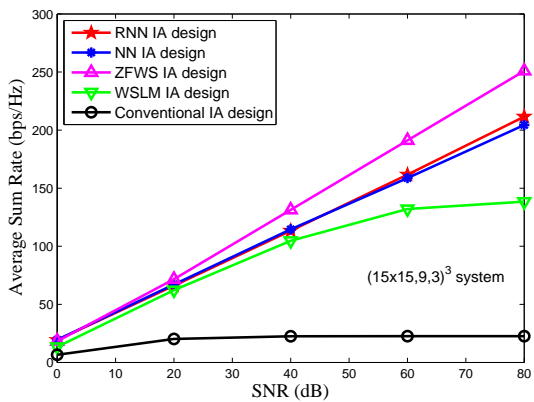}
\vspace{-0.3cm}
\caption{The average SSR versus SNR for ${\left( 15 \times 15,9,3 \right)^3}$ systems.}
\label{fig3}
\end{center}
\end{figure}

The average SSR for a $(15\times 15,9,3)^3$ system is plotted in Fig. \ref{fig3}. It can be observed that our proposed IA algorithms offer the same SSR for SNRs lower than $40$ dB but in high SNRs, the RNN IA design provides a slight SSR improvement as compared with the NN IA design. The reason is that the later uses better surrogates of $\mathrm{rank}$ function than the former. In this system, all IA designs still outperform the conventional IA design, and the proposed IA schemes perform better than the WSLM IA algorithm. However, they perform worse than the ZFWS IA algorithms. The reason is probably that the perfect IA is not guaranteed although the considered system is proper.
\section{Conclusion}
\label{sec:Conclusion}
In this work, we presented the IA schemes for a secure multiuser MIMO system in presence of an eavesdropper. We reformulate the design problem to minimize the rank of interference subspace and wiretapped signal matrices subject to the full rank of the desired signal matrices. To tackle with nonconvexity of rank functions, we introduce two surrogate functions, namely NN and RNN. Then, we developed iterative algorithms based on the coordinate descent approach to obtain suboptimal solutions of the precoding matrices and receive subspace matrices. Numerical results show that the proposed IA designs perform the same or better IA approaches based on minimizing power of interference and wiretapped signals in the system where the perfect IA can be achieved. In addition, both proposed IA algorithms outperform the conventional IA algorithm in terms of SSR.

\bibliographystyle{IEEEtran}
\bibliography{IEEEabrv,newidea2015}
\end{document}